%% file: main.tex
\documentclass[12pt, a4paper]{article}
\usepackage[pdftex]{graphicx}

\input{preamble}

\usepackage{placeins}
\usepackage{subcaption}

\usepackage{xcolor}
\definecolor{mygray}{gray}{0.95}

\usepackage[ruled,vlined]{algorithm2e}
\usepackage{accents}

\usepackage{threeparttable}

\usepackage{todonotes}
\begin{document}

\title{Hamiltonian Monte Carlo for Regression with High-Dimensional Categorical Data\thanks{\textbf{This paper has now been superceeded by "Inference for Regression with Variables Generated from Unstructured Data" (Battaglia, Hansen, Christensen \& Sacher, 2024) (https://arxiv.org/abs/2402.15585)} Hansen gratefully acknowledges financial support from ERC Consolidator Grant 864863.  The authors also thank the NumPyro development team for their outstanding work.}}

\author{
Szymon Sacher \\ Columbia University  \and
Laura Battaglia \\ Oxford University \and
Stephen Hansen \\ University College London
}

\date{\today}
\maketitle

\begin{abstract}

    Latent variable models are increasingly used in economics for high-dimensional categorical data like text and surveys. We demonstrate the effectiveness of Hamiltonian Monte Carlo (HMC) with parallelized automatic differentiation for analyzing such data in a computationally efficient and methodologically sound manner. Our new model, Supervised Topic Model with Covariates, shows that carefully modeling this type of data can have significant implications on conclusions compared to a simpler, frequently used, yet methodologically problematic, two-step approach. A simulation study and revisiting \cite{bandieraCEOBehaviorFirm2020}'s study of executive time use demonstrate these results. The approach accommodates thousands of parameters and doesn't require custom algorithms specific to each model, making it accessible for applied researchers

\end{abstract}

\clearpage{}

\onehalfspacing

\section{Introduction} \label{sec:intro}
\input{intro}

\section{Supervised Topic Model with Covariates} \label{sec:model}
\input{model}

\section{Hamiltonian Monte Carlo} \label{sec:hmc}
\input{hmc}

\section{Results} \label{sec:results}
\input{results}

\section{Conclusion} \label{sec:conclusion}
\input{conclusion}

\newpage

\bibliography{unstruct.bib}

\FloatBarrier
\newpage
\appendix
\numberwithin{table}{section}
\numberwithin{figure}{section}
\input{appendix}
\end{document}

%% file: preamble.tex
\usepackage{lscape}
\usepackage{setspace}
\usepackage{amssymb,amsmath}
\usepackage{amsfonts}
\usepackage{verbatim}
\usepackage{fullpage}
\usepackage{bbm}
\usepackage{natbib}
\usepackage{multirow}
\usepackage{geometry}
\usepackage{hyperref}
\usepackage{color}
\hypersetup{citebordercolor= 1 1 1, linkbordercolor = 0 1 0}
\usepackage{booktabs}
\usepackage{adjustbox}
\bibpunct{(}{)}{,}{a}{}{,}
\usepackage{bm}
\usepackage{textcomp}

\usepackage[labelfont=bf]{caption}
\usepackage{rotating}

\usepackage{tabularx}

%% file: intro.tex
As the amount of digitally recorded unstructured data continues to grow rapidly, empirical work in economics is increasingly incorporating it.  The leading example of such data is text, which numerous papers in a variety of fields have recently used \citep{gentzkowTextData2019}, but also includes others such as survey responses, images, and audio recordings.  The most relevant feature of unstructured data for statistical modeling is that observations typically have an enormous number of independent dimensions of variation.  Moreover, this variation is often expressed in terms of integer counts rather than continuous variables.\footnote{For example, one of the simplest representations of a textual corpus is the \textit{bag-of-words} model in which each document is represented as a vector of integer counts over the unique vocabulary terms in the corpus.  Even relatively small corpora contain thousands of unique dimensions.  Moreover, the dimensionality grows even further as one consider richer linguistic units than individual words.}  Effectively handling such high-dimensional categorical data is therefore a major challenge in extracting information from unstructured data.

One common approach in the literature is to specify a statistical (typically Bayesian) model that projects each observation onto a low-dimensional latent space that captures the important variation in the high-dimensional feature space.  In the natural language context, a popular latent variable model is \textit{latent Dirichlet allocation} \citep[LDA,][]{bleiLatentDirichletAllocation2003}.  In LDA the latent space represents ``topics'' and each document is a mixture over different topics.  A recent selection of applications of LDA include macroeconomic forecasting \citep{larsenValueNewsEconomic2019,bybeeStructureEconomicNews2020,thorsrudWordsAreNew2020,ellingsenNewsMediaVs2021}; conflict forecasting \citep{muellerReadingLinesPrediction2018}; asset pricing \citep{hanleyDynamicInterpretationEmerging2019,lopezliraRiskFactorsThat2019}; political deliberation \citep{hansenTransparencyDeliberationFOMC2018,stiglitzObservabilityReasonedDiscourse2020}; central bank communication \citep{hansenShockingLanguageUnderstanding2016,hansenLongrunInformationEffect2019,dieijenWhatSayThey2019}; corporate finance \citep{adamsDeathCommitteeAnalysis2021}; and media economics \citep{nimarkNewsMediaDelegated2019,bertschNarrativeFragmentationBusiness2021}.  Latent variable models for other data types closely related to LDA have also been adapted for survey data \citep{bandieraCEOBehaviorFirm2020,munroLatentDirichletAnalysis2020,dracaHowPolarizedAre2021} and for network data \citep{nimczikJobMobilityNetworks2017,olivellaDynamicStochasticBlockmodel2021}.

In economics and finance, latent variable models often serve as a preliminary means of transforming unstructured data into a tractable, numeric form that is subsequently utilized in a second-step regression model; this regression model generally treats the transformed information as given data. However, such a procedure presents a few methodological issues\footnote{\citep{egamiHowMakeCausal2018} discuss the difficulties with causal interpretation of results obtained with high-dimensional data such as the potential for over--fitting and identification problems.}. First, uncertainty in the latent representation from the first step is ignored in the second step, which invalidates standard inference procedures. Equally weighting observations may also be inefficient when the individual estimates vary in precision.  Furthermore, the regression model usually imposes dependencies between latent representations and covariates that are ignored in the first step.  This implies a potential loss of information in the first step, as assumptions about the relationship between data and covariates are ignored in the construction of the latent space.

In this paper we show how To overcome these problems in the most natural way -- by jointly specifying, and estimating, the latent variable model and associated regression models within a single, integrated data generating process.  While formulating such integrated models is relatively straightforward, conducting inference for them has to date been anything but. We approach this problem with Hamiltonian Monte Carlo \citep[HMC,][]{mackayInformationTheoryInference2003,nealMCMCUsingHamiltonian2012}, a Markov Chain Monte Carlo (MCMC) algorithm that uses information on the gradient of a joint distribution to sample from it.

To illustrate the approach we introduce a new model, the Structural Topic Model with Covariates and apply it to the data from \cite{bandieraCEOBehaviorFirm2020}.  The authors sought to represent salient differences in executive time use -- an inherently high-dimensional object -- as a low-dimensional and interpretable index, at to correlate this index with CEO and firm characteristics, and firms' performance. Our model is developed to explicitly account for all of these dependencies. Despite the apparent similarity in uncovered patterns of behavior we find substantial differences in the estimates of the relationship between the index and firm performance and CEO characteristics. For instance, while the two-step approach finds that the CEO index is negatively (but statistically insignificantly) correlated with the indicator for CEO obtaining an MBA degree, we find that the MBA has a large, positive and statistically significant effect on the index.

To further demonstrate the usefulness of HMC for estimating structural models with high-dimensional data we also replicate the study of \cite{munroLatentDirichletAnalysis2020}. In this paper the authors developed a custom inference algorithm to study how common patterns in survey responses evolve over time. We show that HMC can be used to estimate the model while improving computational performance and, critically, the quality of inference. What is more, our implementation uses fewer than 25 lines of code. As such, we believe that HMC is a very promising tool for estimating structural models with high-dimensional data. 

One of the main contributions of our paper is to evaluate HMC implemented via parallelized automatic differentiation as a means of conducting inference in latent variable models for high-dimensional data.  To do so, we use the NumPyro package \citep{binghamPyroDeepUniversal2018,phanComposableEffectsFlexible2019}, one of whose key innovations is the efficient computation of the gradients that underlie HMC. It is worth noting that while HMC, as implemented in Stan \cite{carpenterStanProbabilisticProgramming2017} has been previously used in applied Bayesian econometrics \citep[e.g.][]{meagerUnderstandingAverageImpact2019,bandiera_women_2021}, these applications have been limited to simple Bayesian meta-analyses with a few dozen parameters.  In contrast, by using massive parallelization on a dedicated hardware (Graphical Processing Unit, GPU) we are able to estimate models with tens of thousands of parameters in a matter of minutes.  This is a major step forward in the application of HMC to applied Bayesian econometrics.

The rest of the paper proceeds as follows. Sec \ref{sec:model} introduces the Structural Topic Model with Covariates. Section \ref{sec:hmc} provides a brief overview of HMC. Section \ref{sec:results} reports the results of the simulation study, the application to the time use survey of \citet{bandieraCEOBehaviorFirm2020} and the application to the dynamic latent variable model of \citet{munroLatentDirichletAnalysis2020}. Section \ref{sec:conclusion} discusses the results and concludes.

%% file: model.tex
Since its introduction Latent Dirichlet Allocation \citep{bleiLatentDirichletAllocation2003}---perhaps the most well-known topic model--- has seen numerous extensions introducing additional structure. Two of the most popular extensions are the Supervised LDA \citep{bleiSupervisedTopicModels2010} and Structural Topic Models \citep{robertsStructuralTopicModels2014}. The former introduces a numerical response variable, whose distribution is influenced by topic shares. The latter allows for topics to be influenced by a set of covariates, or document-level metadata. 

Below, we propose a model that combines both of these channels and call it Supervised Topic Model with Covariates. In the model, an observation, $x_{i}$, is a $V-$dimensional vector of counts, so $x_{i,j}$ is the number of times a \textit{feature} $j$ appears in observation $i$. In the context of text, $x_{i,j}$ is the number of times word $j$ appears in document $i$, while in the CEO behavior context studied below, $x_{i,j}$ is the number of times CEO $i$ engages in behavior $j$. Each observation is characterized by an unobserved vector of \textit{type shares} $\theta_i$ (often referred to as topics), distribution of which depends on the observed vector of covariates $g_i$. In keeping with literature, we assume that $\theta_i$ follows multivariate Logistic Normal distribution where the location parameters are linear functions of the covariates.\footnote{Note that the effect of a particular covariate on the share of topic $k$ depends only on the differences between $\phi_k$ and all other parameters $\phi_{k'}$ (rather than on the levels). Therefore, we shall normalize $\phi_k$ to zero for one (typically the last) topic.}  

We additionally assume there exists a numerical response variable $y_i$ which is also a linear function composed of topic shares and other control variables, $q_i$. Depending on th application the variables $g_i$ and $q_i$ may or may not be the same. Since, by construction, the type shares sum to one, $\gamma_k$ can be normalized to 0 for a single topic.

As is usual in topic models, the counts in $c_i$ are assumed to be generated from a Multinomial distribution with parameters $\sum_k \theta_{i,k}\beta_k$, where $\beta_k$ is the distribution of the $k$th topic over the $V$ possible features. The model is summarized in the Equation \ref{eq:suptmc_likelihood}. We also present the model graphically as as a plate diagram in Figure \ref{fig:plate-sslda}. What is important, is that the model correctly treats topic shares as unobserved, and therefore the respective the respective node is unshaded. 

\begin{equation} \label{eq:suptmc_likelihood}
    \begin{aligned}
        \theta_i &\sim \text{LogisticNormal}\left[({g}_i^T\phi_1, \ldots, {g}_d^T\phi_K)^T,\textrm{I}\sigma^\theta\right] &(\text{Topic Shares}) \\
        x_{i} &\sim \text{Mult}(N_i, \sum_k \theta_{i,k}\beta_k) &(\text{Categorical Data})\\
        y_i &\sim \text{Normal}(\sum_k\theta_{i,k} \gamma_k + \zeta^Tq_i, \sigma_y^2) &(\text{Numerical Data})\\ 
    \end{aligned}
\end{equation}

\begin{figure}
    \caption{Plate Diagram for the Supervised Topic Model with Covariates.}
    \label{fig:plate-sslda}
    \centering
    \begin{tikzpicture}
        \node[circle, draw, inner sep=0pt, minimum size=1cm, fill=lightgray] (X_i) at (-4, 1) {$ g_i$};
        \node[circle, draw, inner sep=0pt, minimum size=1cm, fill=lightgray] (W_i) at (-4, -2) {$ x_i$};
        \node[circle, draw, inner sep=0pt, minimum size=1cm, fill=lightgray] (Y_i) at (0, -2) {$y_i$};
        \node[circle, draw, inner sep=0pt, minimum size=1cm] (theta_i) at (-2, -0.5) {$\theta_i$};
        \node[circle, draw, inner sep=0pt, minimum size=1cm, fill=lightgray] (V_i) at (-2, -3.5) {$ q_i$};
        
        \node[circle, draw, inner sep=0pt, minimum size=1cm] (beta) at (-6, -2) {$\beta_k$};
        
        \node[circle, draw, inner sep=0pt, minimum size=1cm] (alpha) at (-6, 0) {$\phi_k$};
        
        \node[circle, draw, inner sep=0pt, minimum size=1cm] (phi) at (1.8, 0) {$\gamma_k$};
        
        \node[circle, draw, inner sep=0pt, minimum size=1cm] (zeta) at (1.8, -2) {$ \zeta_q$};
        
        \node[circle, draw, inner sep=0pt, minimum size=1cm] (sigmay) at (1.8, -4) {$\sigma_{y}$};

        \draw[->, >=stealth] (X_i.south) -- (theta_i.north);
        
        \draw[->, >=stealth] (beta.east) -- (W_i.west);
        \draw[->, >=stealth] (alpha.east) -- (theta_i.west);
        
        \draw[->, >=stealth] (theta_i.south) -- (W_i.north);
        \draw[->, >=stealth] (theta_i.south) -- (Y_i.north);
        \draw[->, >=stealth] (V_i.north) -- (Y_i.south);
        
        \draw[->, >=stealth] (zeta.west) -- (Y_i.east);
        \draw[->, >=stealth] (phi.west) -- (Y_i.east);
        \draw[->, >=stealth] (sigmay.west) -- (Y_i.east);

        \draw (-6.8, -3) rectangle (-5.2, -1.2);
        \node at (-5.4, -2.7) {$K$};
        
        \draw (-6.8, -1) rectangle (-5.2, .8);
        \node at (-5.4, -.7) {$K$};
        
        \draw (1, -1) rectangle (2.6, .8);
        \node at (2.4, -.7) {$K$};
        
        \draw (1, -3) rectangle (2.6, -1.2);
        \node at (2.3, -2.7) {$| q_i|$};
        
        \draw (-4.8, -4.3) rectangle (0.8, 1.8);
        \node at (.4, -4.0) {$D$};
    \end{tikzpicture}

    \footnotesize
    \textbf{Note:} The figure presents a simplified diagram of two periods of the Supervised Topic Model with Covariates. The number of topics/types $K$ is a hyperparameter to be specified. The rectangles display conditionally independent observations and latent variables. The shaded circles represent observed data, while the white circles represent (unobserved) latent variables. Prior distributions on latent variables are omitted for simplicity.

\end{figure}
To finish, the prior distributions need to be defined. We chose to use the Dirichlet distribution for $\beta_k$, independent Normal distribution for each of the regression parameters, and the Gamma distribution for the standard deviations of the regression errors. These are all outlined in Equation \ref{eq:suptmc_priors}. These choices are consistent with prior literature. Importantly, however the inference algorithm explained in Section \ref{sec:hmc} does not require conjugate priors, and so it is possible to change them if needed with just one line of code.

\begin{equation} \label{eq:suptmc_priors}
    \begin{aligned}
        &\bm\beta_k \sim \text{Dirichlet}(\eta) \\
        &\phi_{k} \sim \text{Normal}(0,\sigma^\phi) \\
        &\gamma_k \sim \text{Normal}(0, \sigma^\gamma) \\
        &\zeta_q \sim \text{Normal}(0, \sigma^\zeta) \\
        &\sigma_y \sim \text{Gamma}(s_0,s_1)
    \end{aligned}
\end{equation}

Before describing the inference method we use, we want to reiterate that the model presented above serves primarily as an illustration. An interested researcher should be able to modify it as needed to accommodate different data, to test robustness of the conclusions to specifying alternative distributions for the data, or with respect to choice of priors. In that, researcher is not limited to specifying only models that can be conveniently estimated (for example by exploiting conjugacy as in the Gibbs sampler). As a further illustration of the usefulness and flexibility of HMC as a general-purpose inference algorithm, in Section \ref{sec:dynamicModel} we report replication of a dynamic model of \citet{munroLatentDirichletAnalysis2020}, where the topic distributions are allowed to evolve smoothly over time.

%% file: hmc.tex
Presenting an in-depth of Hamiltonian Monte Carlo algorithm is beyond the scope of the paper, and our goal here is to give a high-level overview and an illustration of HMC applied to the model presented in Section \ref{sec:model}. Excellent articles that cover the basic ideas that underpin HMC are \cite{nealMCMCUsingHamiltonian2012}, \cite{hoffmanNoUTurnSamplerAdaptively2014} and \cite{betancourtConceptualIntroductionHamiltonian2018}. We are not aware of the application of HMC to LDA or related models in the literature.

The HMC algorithm is a variant of a popular Metropolis-Hastings (MH) algorithm, a well-known Markov chain Monte Carlo (MCMC) method that generates samples from a target distribution in two steps: (1) propose a new state $\phi'$ from the current state $\phi$ using a pre-specified proposal distribution; (2) accept the new proposal with a probability that depends on the relative posterior density at the new proposal and the old sample (proposals with a higher joint density are accepted more often). The MH is a general-purpose inference algorithm that can be applied to any target distribution, as long as the joint density of data $x$ and parameters $\phi$, $q(x, \phi)$, can be evaluated. The most common variant of MH, the Random Walk Metropolis-Hastings, proposes new states according to a random walk. This, however often leads to problems with convergence. The random walk behavior means that, in order to maintain reasonable acceptance probability, especially in high dimensions, the steps taken have to be very small as taking a large step in a random direction can drastically reduce the joint density. This, in turn, results in a very high autocorrelation in samples and prohibitively slow convergence. The problem is particularly severe if the parameter space is high-dimensional and the target distribution cannot be well approximated by a product of marginal densities.  

The HMC algorithm and its variants utilize the geometry of joint density to propose distant states that nonetheless have high chance of acceptance. This is achieved by proposing a new state $\phi'$ by following Hamiltonian dynamics for a certain number of steps, starting from the initial state $\phi$. This process is determined by the curvature of the joint density, so to approximate it, it is necessary to evaluate the gradient $\frac{d{q}}{d\phi}$ of the joint density with respect to the parameters $\phi$. In practice, modern probabilistic programming libraries use automatic differentiation to compute the gradients for an arbitrarily-defined density. Furthermore, the densities and gradients are typically computable in a parallel manner due to them being additive with respect to the data points.

The specific variant of HMC that we use is the No-U-Turn Sampler (NUTS, \cite{hoffmanNoUTurnSamplerAdaptively2014}) implemented in NumPyro, a library for Python \citep{phanComposableEffectsFlexible2019}. The intuitive idea of NUTS is to follow Hamiltonian dynamics until the resulting path begins to circle back to its starting point. This is computationally advantageous since it generates proposals relatively far from each other, thus reducing correlation between draws, and reduces the number of steps needed to generate a proposal. The NUTS algorithm is also adaptive, which means that it automatically tunes the step size to achieve a desired acceptance rate. The No-U-Turn Sampler is implemented in many probabilistic programming libraries. The main advantage of NumPyro relative to Stan---a much more popular implementation--- is that it utilize a state-of-the-art automatic differentiation engine, Jax \citep{jax2018github} and allows users to deploy these computations to specialized hardware such as Graphical Processing Units (GPUs) and Tensor Processing Units (TPUs), resulting in a dramatic improvement in computation time. What is more, Numpyro is a Python library, not a standalone program, which means that is easy to integrate with other libraries and benefits from host of functionalities that Python provides.

%% file: results.tex
\subsection{Simulation} \label{subsec:simulation}

We start by reporting the results of a simple simulation exercise designed to illustrate the shortcomings of the typical, 2-step approach to estimating such models. In that approach a researcher first estimate the type shares using an unsupervised, often black box algorithm (such as those found in Python's \textit{Gensim} package), and then uses resulting point estimates as data in down-stream regressions. We contrast the 2-step approach with the HMC implementation which estimates the model jointly.

We simulated the data according to the data generating process described in Equations \ref{eq:suptmc_likelihood} and \ref{eq:suptmc_priors}; further details are included in Appendix Section \ref{sec:appendix_simulation}. In each of the 50 simulations there are $I = 1000$ observations of length $N_i = 200$ with $J=500$ features. For simplicity we use $K=2$ types and focus on two coefficients: (1) $\gamma_1$, the effect of the increase in (unobserved) share of Type 1 on numerical outcome; (2)$\phi_1$, the effect of a numerical covariate on the (un-normalized, real-valued) Type 1 share. 

\begin{figure}
     \centering
     \begin{subfigure}[b]{0.45\textwidth}
         \centering
         \subcaption{Estimates: 2-step approach}
         \label{fig:gamma_2steps_coefs}
         \includegraphics[width=\textwidth]{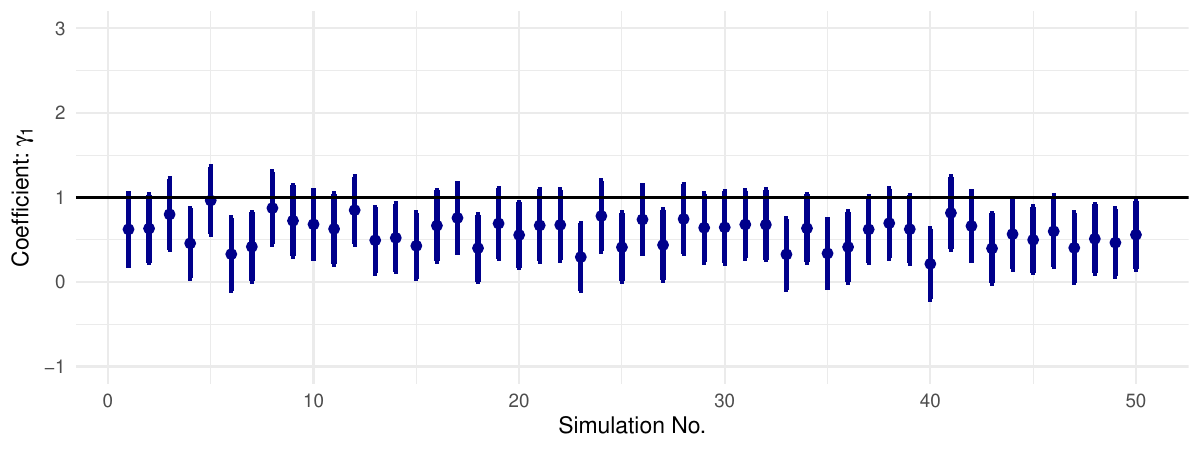}
     \end{subfigure}
     \hfill
     \begin{subfigure}[b]{0.45\textwidth}
         \centering
         \subcaption{Estimates: HMC approach}
         \label{fig:gamma_hmc_coefs}
         \includegraphics[width=\textwidth]{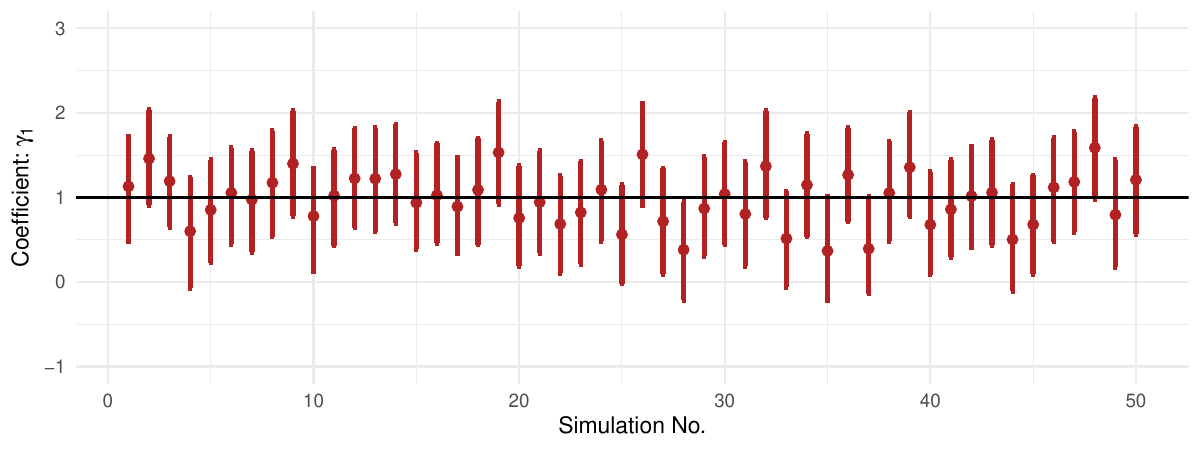}
     \end{subfigure}
     \hfill
     \begin{subfigure}[b]{\textwidth}
         \centering
         \subcaption{Frequency of inclusion in interval estimate}
         \label{fig:gamma_hmc_2steps_props}
         \includegraphics[width=0.6\textwidth]{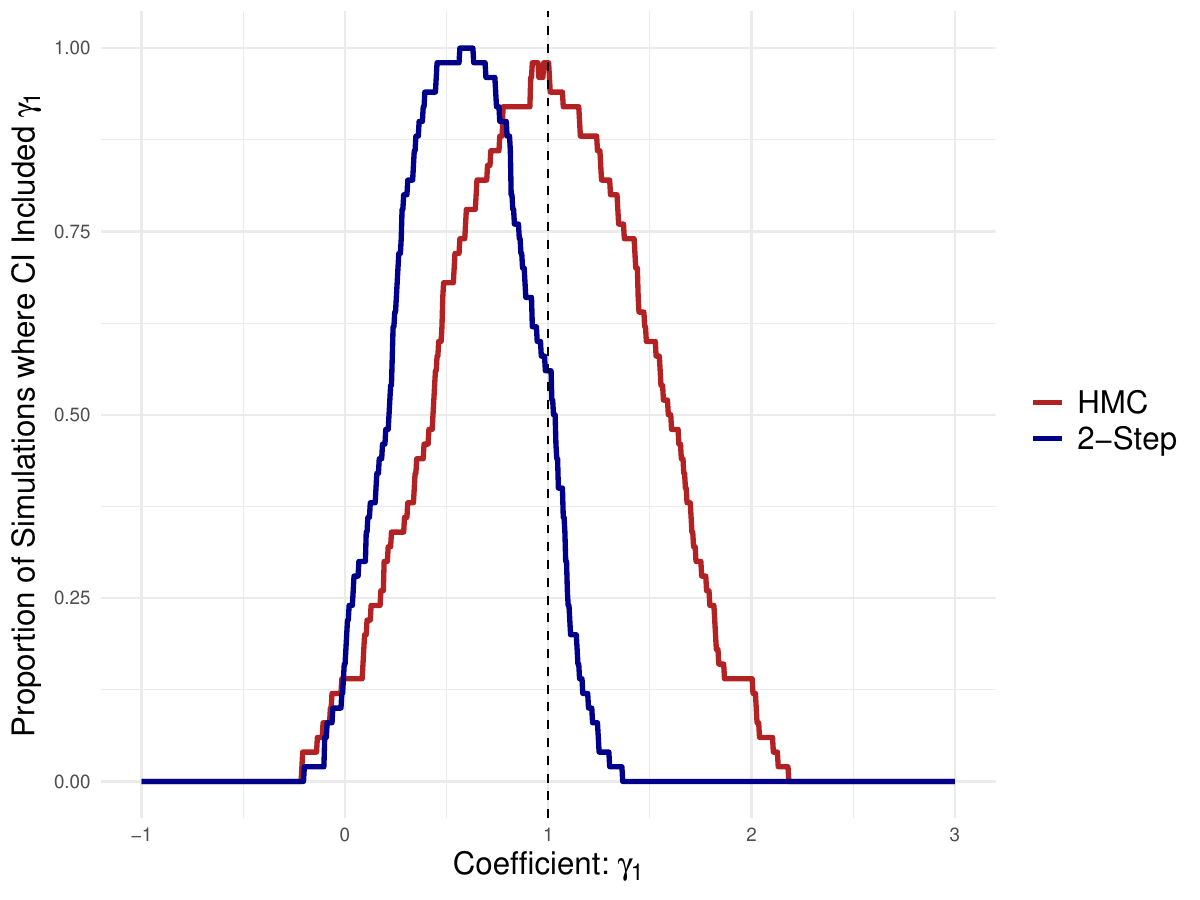}
     \end{subfigure}
     \caption{Simulation Results: Effect of Type Shares on Outcomes \label{fig:gamma_sim_results}}  
        \begin{quote}
            \footnotesize
            \textbf{Note:} The figures show results from the simulation exercise comparing three inference algorithms for estimating the $\gamma_1$ coefficient in a Supervised Topic Model with Covariates, which captures the impact of variation in a covariate on class shares. The true value of $\gamma_1$ is $1$.  The algorithms were used to estimate the coefficient across 50 simulations.  The top panel shows point estimates as dots and interval estimates as vertical lines around the dots.  The horizontal grey line is the true value.  The bottom figure displays the fraction of simulations in which different values of $\gamma_1$ appear in interval estimates. Noticeably, the estimates using 2-step approach are biased down and their standard errors are under-estimated.  
        \end{quote}

\end{figure}

Figure \ref{fig:gamma_sim_results} presents the results for $\gamma_1$. Panel \subref{fig:gamma_2steps_coefs} demonstrates that the point estimates from the 2-step approach are biased towards zero and the standard errors are under-estimated. This illustrates two issues with the 2-step approach. First, since in this regression the inferred type share is used as an independent variable, a mis-measurement in the type shares results in attenuation bias. Second, these estimates are treated as data, without accounting for the uncertainty, and as a result the standard errors are underestimated. On the other hand, the HMC approach performs much better. Panel \subref{fig:gamma_hmc_coefs} shows that point estimates are distributed around the true value, and the bayesian credible intervals have the expected coverage -- the power is not artificially inflated. The true value of $\gamma_1$ is included in all but one of the credible intervals, resulting in a slightly larger coverage than the nominal 95\% level. 

\begin{figure}
    \centering
     
    \begin{subfigure}[b]{0.45\textwidth}
        \centering
        \caption{Estimates: 2-step approach}
        \label{fig:phi_2steps_coefs}
        \includegraphics[width=\textwidth]{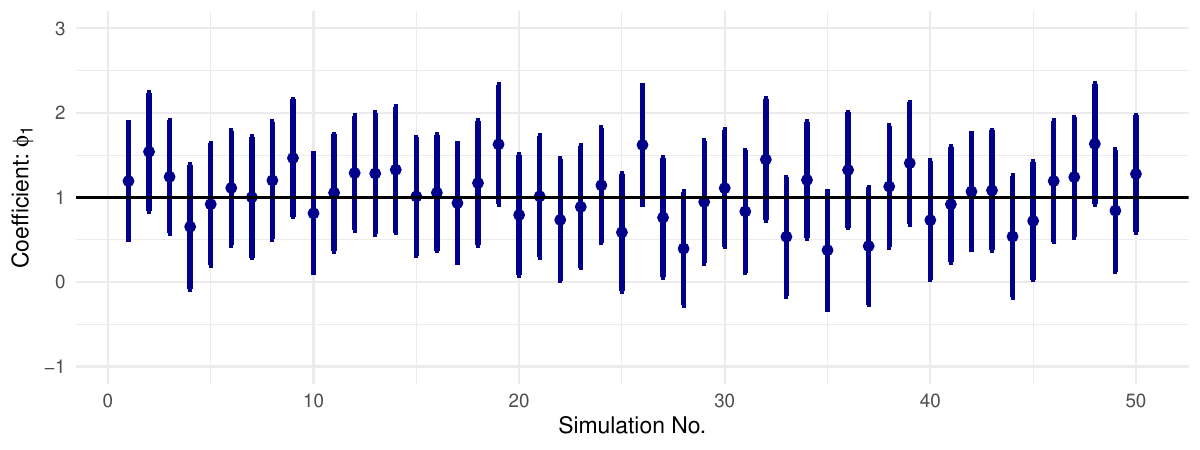}
    \end{subfigure}
    \hfill
    \begin{subfigure}[b]{0.45\textwidth}
        \centering
        \caption{Estimates: HMC approach}
        \label{fig:phi_hmc_coefs}
        \includegraphics[width=\textwidth]{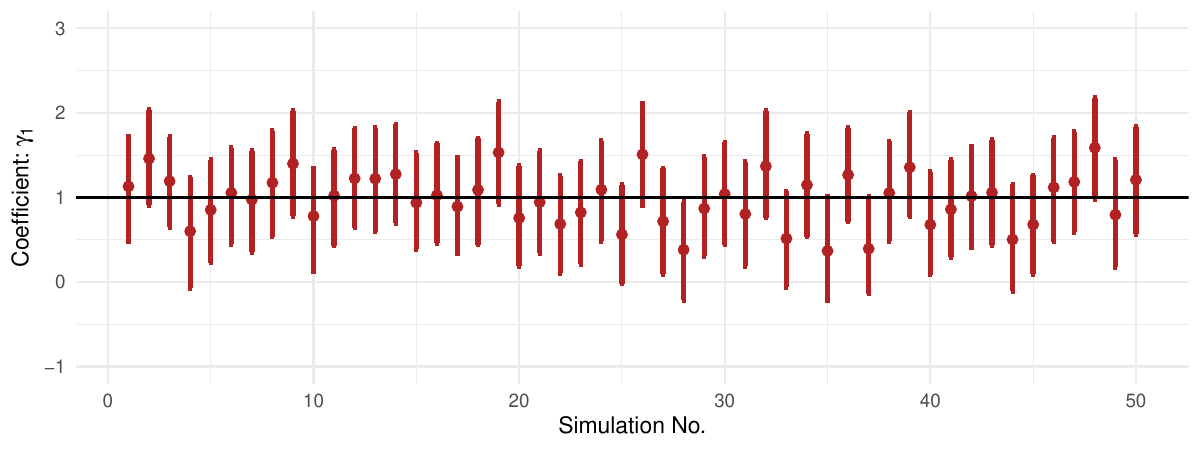}
    \end{subfigure}
    \hfill
    \begin{subfigure}[b]{\textwidth}
        \centering
        \caption{Frequency of inclusion in interval estimate}
        \label{fig:phi_hmc_2steps_props}
        \includegraphics[width=0.6\textwidth]{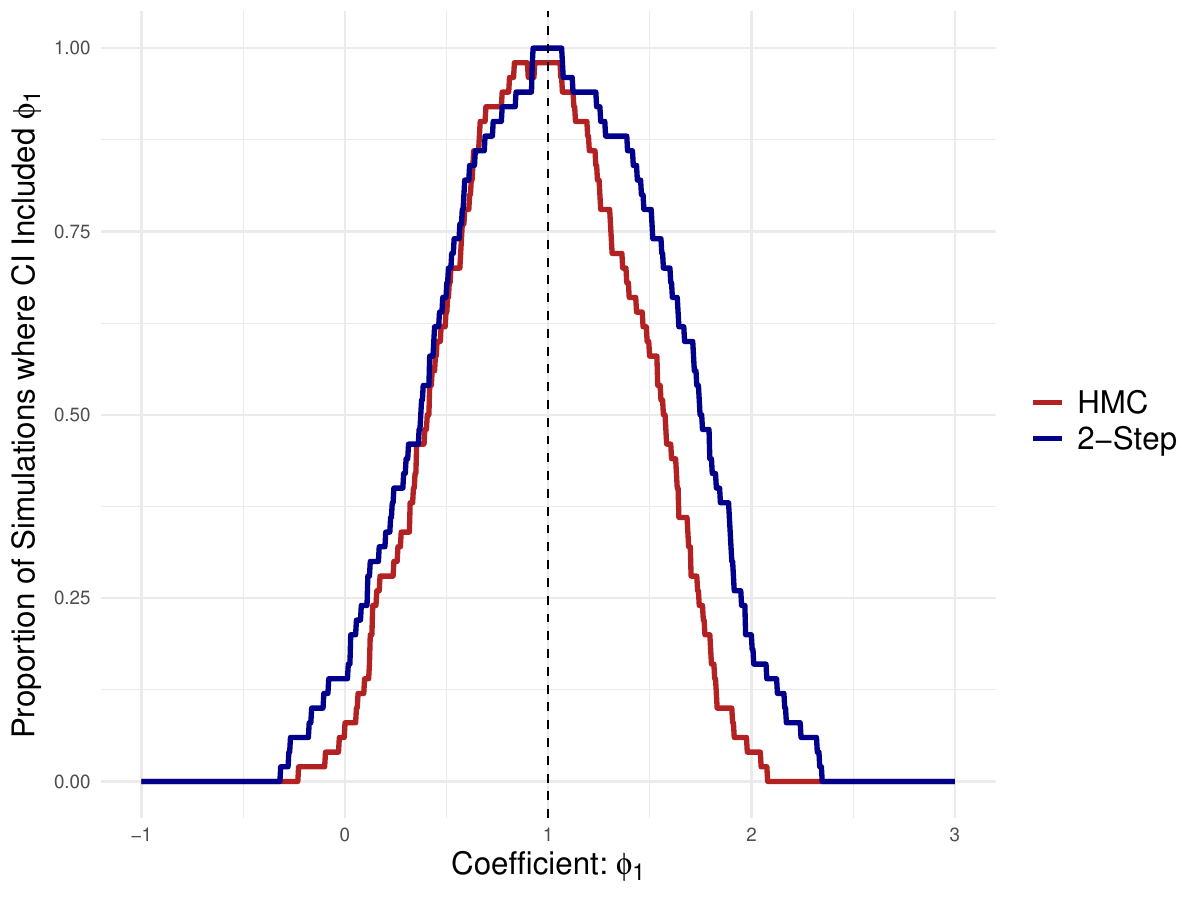}
    \end{subfigure}
    \caption{Simulation Results: Effect of a Covariate on Type Shares \label{fig:phi_sim_results}}   
       \begin{quote}
           \footnotesize
           \textbf{Note:} The figures show results from the simulation exercise comparing three inference algorithms for estimating the $\phi_1$ coefficient in a Supervised Topic Model with Covariates, which captures the impact of variation in a covariate on class shares. The true value of $\phi_1$ is $1$.  The algorithms were used to estimate the coefficient across 50 simulations.  The top panel shows point estimates as dots and interval estimates as vertical lines around the dots.  The horizontal grey line is the true value.  The bottom figure displays the fraction of simulations in which different values of $\phi_1$ appear in interval estimates. Both methods are unbiased, however the 2-step approach has lower power. 
       \end{quote}
\end{figure}

Figure \ref{fig:phi_sim_results} presents the results for $\phi_1$, the effect of the covariate on the type shares. Notice, in this case the estimated latent variables, $\theta$, are treated as the dependent variable therefore the attenuation bias mentioned above is not present and both approaches are approximately unbiased. However, in this case, the potential mis-measurement of $\theta_i$ results in a lower power for the 2-step approach, as seen in Panel \subref{fig:phi_hmc_2steps_props}.\footnote{Separately, there is an issue of in-efficient weighting in the 2-step approach. We discuss it further in the next subsection.}   

Finally, a word on the computational performance is in order. We have found that Numpyro's HMC implementation of the model is fast -- each simulation took approximately 3 minutes and was comparable to the 2-step approach implemented with Python's \textit{lda} package that uses the collapsed Gibbs sampler of \cite{griffithsFindingScientificTopics2004}. Further, as expected we found that the HMC implementation benefits significantly from GPU acceleration. For instance, on the GPU we were able to perform, on average, 2742 gradient evaluations per second, compared with approximately 260 on a CPU. Consequently, the GPU implementation run approximately 10 times faster than CPU one. Further details are available in Appendix \ref{sec:appendix_simulation}.

\subsection{Application: CEO Behavior}
To show that modelling and jointly estimating the internal structure of data matters in practice, we revisit the study of \citet{bandieraCEOBehaviorFirm2020}, who collected data on CEO's activities with the goal to describe salient differences in executive time use, and relate those to firm and CEO characteristics as well as firm outcomes. 

A cross section of 916 CEOs participated in a survey that recorded features of time use in each 15-minute interval of a given week, e.g. Monday 8am-8:15am, Monday 8:15am-8:30am, and so forth.  The recorded categories are 1) the type of activity (meeting, public event, etc.); 2) duration of activity (15m, 30m, etc.); 3) whether the activity is planned or unplanned; 4) the number of participants in the activity; 5) the functions of the participants in the activity (HR, finance, suppliers, etc.). In total there are 654 unique combinations of these categories observed in the data, which we will refer to as feature $j$. We can denote $x_{i,j}$ as the number of times feature $j$ appears in the time use diary of CEO $i$.  On average a CEO is engaged in 88.4 activities, with a minimum of 2 and a maximum of 222.

\citet{bandieraCEOBehaviorFirm2020} use LDA with $K=2$ dimensions to organize the time use data.  The authors refer to these dimensions as \textit{pure behaviors}, and each one gives a separate distribution over time use combinations $\beta_1$ and $\beta_2$. The share of CEO's time devoted to pure behavior $1$,  $\theta_{i,1}$, is referred to as the \textit{CEO index}.

The authors first estimate LDA on the time use data using the collapsed Gibbs sampler of \citet{griffithsFindingScientificTopics2004}, then form an estimate $\hat{\theta}_{d,1}$ based on the posterior means.  They then use $\hat{\theta}_{d,1}$ as an input into the productivity regression where $y_i$ is the log of firm $i$ sales, $q_{i}$ is a vector of firm observables. Further, they separately analyzed which CEO and firm characteristics are associated with behaviors. This question was investigated by regressing the estimated $\hat{\theta}_{d,1}$ on a vector of characteristics, $g_i$.

We reexamine these questions by applying Supervised Topic Model with Covariates to this data. In both $g_i$ and $q_i$ we include log of firm's employment and a list of controls that authors deemed important (such as country and year). To explain CEO behavior in $g_i$ we also add an indicator for whether or not the CEO has an MBA. For comparison, in addition to the results obtained with Sup-TM-C, we also report the results obtained with the original authors' 2-step (separated) approach, and those where we do not use any of the covariates $g_i$.

\begin{table}[htb]
    \centering
    \caption{Comparison of types (Pure Behaviors)}
    \begin{threeparttable}
    \input{Figures/StructuralCEO/comparisons.tex}

    \begin{tablenotes}
    \footnotesize
    \item \textbf{Note}:
    This table reports the relative probability of observing certain activities between Pure Behavior 1 and Pure Behavior 2. The value of 1 indicates that this activity is equally likely under both Pure Behaviors. Values higher than 1 mean that this type of activity is more likely to be performed under Pure Behavior 1. The values are reported in columns (1) and (2) are computed by first obtaining mean posterior probabilities of each activity in the given types (Pure Behavior). In column (3) we report values presented in \cite{bandieraCEOBehaviorFirm2020}.
    \end{tablenotes}
    \end{threeparttable}
    \label{tab_pureBehaviorsComparisons}
\end{table}

Turning to results, in Table \ref{tab_pureBehaviorsComparisons} we show that estimated pure behaviors are qualitatively very similar. All three approach suggest that interacting with C-Suite, spending time communicating, and holding multi-function meetings are much more likely under Pure Behavior 1, while spending time on plant visits, and interacting solely with suppliers are more likely under Pure Behavior 2. Based on these observations, the original authors label the CEO's with high values of $\hat{\theta}_{i,1}$ as \textit{leaders} and those with low values as \textit{managers}.

\begin{table}[ht]
    \centering
    \caption{Model Coefficients.}
    \begin{adjustbox}{max width=\textwidth}
        \begin{threeparttable}
            \input{Figures/StructuralCEO/coefTable.tex}
            \begin{tablenotes}
                \footnotesize
                \item \textbf{Note}: In columns (2), (3), and (5) the point estimates are the mean posterior values and the intervals report $95\%$ (symmetric) Bayesian credible intervals. Column (1) and (4) report frequentist estimates and confidence intervals. The Un-normalized CEO index is a real valued variable. In order to obtain the CEO Index from the Un-normalized CEO Index one needs to apply the softmax transform. Conversely, to estimate the coefficient on the Un-normalized index, in column (4) we use a generalized linear model (`glm' in R). MBA is a dummy indicating if a CEO has MBA degree.
            \end{tablenotes}    
        \end{threeparttable}
    \end{adjustbox}
    \label{tab:StructuralCEO-coefs}
\end{table}

Despite the apparent similarity in inferred pure behaviors, we find large differences in regression parameter estimates, as reported in Table \ref{tab:StructuralCEO-coefs}. Panel (a) shows that both the point estimate and the uncertainty around the effect of the CEO index on firm productivity appears significantly larger when we use any of the joint approaches in columns (2) and (3), compared to the two-step approach in column (1). This is consistent with the simulation evidence in Section \ref{subsec:simulation}. In particular, the point estimate is 0.083 when we use the original authors' 2-step approach, and 0.34 when we use Sup-TM-C\footnote{Note that one should remain careful in interpreting these results as causal. One possible issue may be over-fitting. For a discussion on have to perform causal inference using high-dimensional data see \cite{egamiHowMakeCausal2018}. }.

In Panel (b) we show that depending on the approach taken, one can come to very different conclusions regarding the effect of MBA education on CEO behavior. In column (4) we see using the 2-step approach we find negative but not statistically significant relationship. Meanwhile, in column (5) we see that using Sup-TM-C we find a positive, large and statistically significant (in a Bayesian sense) relationship. The contrast is likely cause by the difference in how the two approaches weight observations. As mentioned above, some CEOs are engaged in much larger number of activities than others and, consequently, the posterior distribution of their CEO index $\theta_i$ is much more concentrated around the mean.\footnote{The posterior variance depends not only on the number of activities but also on what particular activities a CEO does. Some activities -- those that are equally likely under both Pure Behaviors -- are much less informative of the CEO index than others and, therefore, does not substantially reduce the posterior variance.}. The regression part of the two-step model ignores this information as it weights all observations equally. In contrast, the joint model correctly incorporates the information about posterior variance and implicitly puts more weight on those observations where the posterior variance is smaller. The differences presented in this section illustrate that the concerns with the typical 2-step approach are not only theoretical but can  have practical implications.

\subsection{Extension: Dynamic Model} \label{sec:dynamicModel}

In the sections above we considered data that can be represented as simple vector of counts. Unfortunately, this representation, is typically not possible when respondents are asked several distinct questions in a survey. The number of possible combinations of answers to these questions grows exponentially fast\footnote{For instance, a survey with $J$ true-false questions has $2^J$ possible combinations of answers.}, however groups of respondents often share similar patterns of answers. For instance, one would expect an `agree' response to the question `do you support abortion rights?' to be positively correlated with an `agree' response to the question `do you support gay marriage?' Uncovering such patterns can be useful to categorize respondents into groups that share similar opinions.

With this goal in mind, \cite{munroLatentDirichletAnalysis2020} propose a model that uses dynamically evolving latent variables (similar to type shares, $\theta$) to uncover both, the common patterns of replies to a survey and shares of respondents who display a given pattern at a given time. Concretely, suppose that in a period $t$ there are $N_t$ respondents to a survey with $J$ questions; that is, the structure of the survey remains the same over time but the size of the population of respondents potentially varies. The parameters $\left\{\boldsymbol\beta^j\right\}_{k = 1}^K$ are $K$ separate distributions that represent response profiles for question $j$ associated with different latent types.  Let $t(i)$ be the time period in which person (i) is observed. Then person $i$ is assigned a type $z_{i,t(i)} \in \{1,\ldots,K\}$ and her response to question $j$, which we denote $x_{i,j,t(i)}\in \{1,\ldots,L_j\}$, is drawn from $\boldsymbol\beta_{z_{i,t(i)}}^j$. The respondent's type, $z_{i,t(i)}$, is drawn from a distribution parametrized by $\theta_t(i)$. In the Appendix \ref{sec:appendix_munrong} we discuss further details of the model including the specific data generating process and priors used.

The original authors implement their model using a custom inference algorithm that combines a Gibbs sampler with a step obtained with help of Langevin dynamics \citep[see]{nealMCMCUsingHamiltonian2012}. In this section we implement \cite{munroLatentDirichletAnalysis2020} model using HMC and replicate their main findings regarding the evolution of `sentiment' in the Michigan Index of Consumer Sentiment survey. Our approach does not require any model-specific derivation and, notably, fewer than 25 lines of code are necessary to implement it. This highlights the usefulness of HMC, as implemented in Numpyro, as an easy-to-use general-purpose inference algorithm.

\begin{figure}
    \centering
    \caption{Type shares in Michigan Index of Survey of Consumers data on consumer confidence.}
    \includegraphics[width=0.9\textwidth]{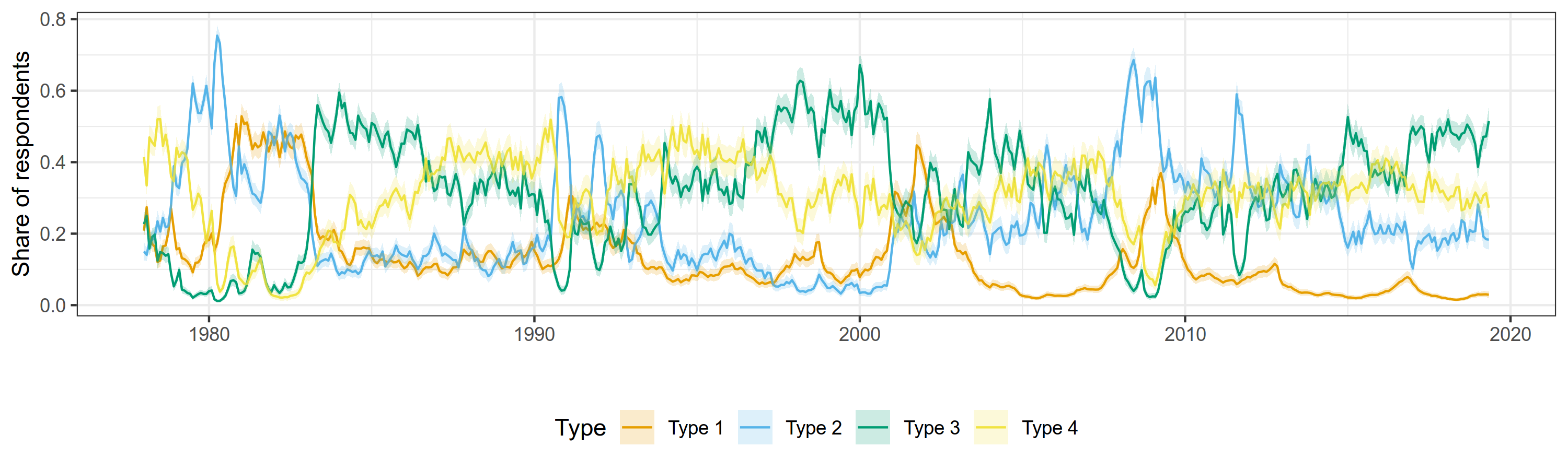}
    \begin{quote}
        \footnotesize
        \textbf{Note:} This figure replicates empirical findings of \cite{munroLatentDirichletAnalysis2020} and plots the inferred type shares $\theta_{k,t}$ estimated on Michigan Consumer Survey data from 1978 through 2019. The solid lines display the posterior means and the shaded areas display $95\%$ (symmetric) Bayesian credible intervals. The authors show that Type 3 closely tracks the Index of Consumer Sentiment and that Type 1 is a leading indicator or recession recovery. 
    \end{quote}
    \label{fig:dynamicModel}
\end{figure}

We use the same data as \cite{munroLatentDirichletAnalysis2020} and report the results in Figure \ref{fig:dynamicModel}. The figure shows the evolution of the inferred type shares, $\theta_{k,t}$ obtained by replicating the survey expectation model of \cite{munroLatentDirichletAnalysis2020}. The solid lines display the posterior means and the shaded areas display $95\%$ (symmetric) Bayesian credible intervals. The evolution of these shares is similar to original finding. The authors show that Type 3 closely tracks the Index of Consumer Sentiment and that Type 1 is a leading indicator or recession recovery (as measured by unemployment rate).

While we find relatively similar point (mean posterior) estimates, our results differ substantially in the inferred posterior variance. Specifically, we find credible intervals that are much wider than in the original paper. This appears to be a consequence of the fact that we are able to achieve much higher effective sample size (ESS) (despite much shorter runtime), indicating that HMC is able to explore the posterior distribution more efficiently that the original algorithm. 

%% file: Figures/StructuralCEO/comparisons.tex
\begin{tabular}{lrrr}
\toprule
              Activity &  Sup-TM &  Sup-TM-C &  Bandiera et al (2020) \\
              & (1) & (2) & (3) \\

\midrule
          Plant Visits &    0.09 &      0.10 &                   0.11 \\
             Suppliers &    0.32 &      0.38 &                   0.32 \\
            Production &    0.40 &      0.41 &                   0.46 \\
        Just Outsiders &    0.72 &      0.74 &                   0.58 \\
         Communication &    1.70 &      1.51 &                   1.49 \\
        Multi-Function &    1.32 &      1.41 &                   1.90 \\
Insiders and Outsiders &    1.89 &      1.97 &                   1.90 \\
               C-suite &   21.69 &     19.88 &                  33.90 \\
\bottomrule
\end{tabular}

%% file: Figures/StructuralCEO/coefTable.tex
\begin{tabular}{lccc|cc} \toprule 
& \multicolumn{5}{c}{Dependent Variable} \\ 
 \cmidrule{2-6} 
& \multicolumn{3}{c}{(a) Log(sales)} & \multicolumn{2}{c}{(b) Un-normalized CEO Index} \\ 
  \cmidrule(lr){2-4} \cmidrule(lr){5-6} 
 & (1) 2-Step & (2) Sup-TM & (3) Sup-TM-C & (4) 2-step & (5)  Sup-TM-C \\ 
 \midrule 
 CEO Index &  0.083 & 0.334 & 0.34  & & \\ 
 &  ( -0.015, 0.181) & (0.179, 0.479)  &  (0.154, 0.516)  & & \\ 
Log Employment &  1.035 & 0.944 & 0.943 & 0.285 & 0.427  \\ 
 &  (0.98992, 1.08008) & (0.903, 0.985) & (0.891, 0.99) & (0.11252, 0.45748) & (0.369, 0.484)  \\ 
MBA & & & & -0.207 & 0.697  \\ 
 & & & & (-0.4716, 0.0576) & (0.542, 0.843)  \\ 
\midrule 

Controls & X & X & X & X & X \\ \bottomrule

 \end{tabular}

%% file: conclusion.tex
In this paper, we developed Supervised Topic Models with Covariates. We applied this novel model to re-examine the study conducted by \cite{bandieraCEOBehaviorFirm2020}. Our analysis demonstrated that if a researcher suspects that there are relationships between latent variables and numerical covariates, then carefully incorporating those into the model can significantly influence the conclusions when compared to simpler but methodologically problematic two-step approaches.

We have also shown how Hamiltonian Monte Carlo deployed with efficient algorithms for automatic differentiation can be used to sample from complex posterior distributions that arise in unstructured data analysis. The results showcase the benefits of this framework when a researcher wants to jointly specify a latent variable model for dimensionality reduction along with a regression model involving those latent representations. Given that most applications of latent variable models for unstructured data in economics fall within this category, we believe our findings will be of wide interest. Applied researchers can now establish problem-specific dependencies and perform valid inference without having to derive and code intricate algorithms. Thus, we expect Hamiltonian Monte Carlo to become an essential component of unstructured data analysis moving forward.

It is also important to acknowledge the limitations of HMC for unstructured data modeling.  One constraint is scalability.  The applications we have explored in this paper do not involve vast amounts of data, nor do many in the literature. The current study and most other applications do not involve massive amounts of data, but for those that do, HMC-based inference may be computationally infeasible.  Historically variational inference algorithms have been used for posterior approximation with large data, and these too can be formulated as probabilistic programs that rely on automatic differentiation \citep{hoffmanStochasticVariationalInference2013}. Recently, scalable extensions of HMC that are based on stochastic approximations of gradients have shown initial promise \citep{dangHamiltonianMonteCarlo2019}. We believe that formulating these procedures as probabilistic programs will be key to their widespread adoption. We leave for future research the question of which inference procedures best suit different contextual conditions of big-data problems.

Identification in LDA and related models is an active area of research, and the choice of priors over categorical distributions can have significant effects on inference even asymptotically \citep{keRobustMachineLearning2021}. These issues, however, are more closely related to the structure of the model than to a specific inference approach. Furthermore, by enabling researchers to focus more on modeling and less on algorithm implementation, the methods we present may aid in empirically evaluating the impact of various prior selections on parameter estimation.

Finally, at a high level, Hamiltonian Monte Carlo (HMC) is an algorithm that is capable of accurately sampling from joint distributions. In economics, structural models are generally expressed as joint distributions over model parameters and data. Therefore, HMC can be utilized for efficient structural estimation. The use of HMC to incorporate unstructured data into structural models holds promising future prospects and paves the way for further developments in the field.

%% file: appendix.tex
\section{Further Details on the Simulation Excercise} \label{sec:appendix_simulation}

Table \ref{app_tab:simulation_params} presents the parameters used for simulation excercise. Since we used $K=2$ classes and class shares must add to 1, only the differences in regression parameters, e.g. $\gamma_1 - \gamma_2$ are identified, therefore in the simulation and estimation we normalized $\gamma_2$ and $\phi_2$ to 0. Second, since class `labels' are not identified in estimation, it is necessary to adjust signs post estimation.

\begin{table} 
    \centering
        \caption{Parameters for the simulation exercise. \label{app_tab:simulation_params}}
        \begin{tabular}{lcc}
            \toprule
            Parameter & Value & Description \\
            \midrule
            $D$ & $1000$ & Number of observations \\
            $V$ & $500$ & Number of distinct features \\
            $N$ & $200$ & Total number of features per document  \\
            $K$ & $2$ & Number of latent types \\
            True $\gamma$ & $(1, 0)$ & Effect of types on numerical outcomes \\
            True $\phi$ & $(1, 0)$ & Effect of a covariate on un-normalized type shares \\
            $Q_i$ & $\sim N(0, \text{log}(3)/1.96)$ & Covariate affecting type shares\\
            $Z_{z, i} \forall {z \in (1,2,3)}$ & $\sim N(0, 3)$ & Additional covariates affecting outcome \\
            $\sigma$ & $4$ & Sd of the outcome's residual \\
            $p(\phi_1)$ & $N(0,2)$ & Prior for $\phi_1$ \\
            $p(\gamma_1)$ & $N(0,2)$ & Prior for $\gamma_1$ \\
            $\eta$ & $0.2$ & Dirichlet concentration parameter \\ 
            \bottomrule 
        \end{tabular}
\end{table}

We performed the simulation on a `N1-highmem-2' instance on the Google Could Platform. The instance has 2 vCPUs and 13 GB of memory. We also utilized a single Tesla T4 GPU. In the 2-step procedure, we run the topic model using Python's `lda' package for 5000 iterations. When using HMC we chose 1000 warmup ans 2000 post-warmup iterations. A single simulation took approximately 1 minute to run using the 2-step procedure and approximately 3 minutes using HMC. Exact runtime will depend on the number of cores and GPU available but we want to highlight that whenever MCMC method such as the Gibbs sampler is used, the HMC should be viable. 

\section{Further Details on \cite{munroLatentDirichletAnalysis2020} Model}
\label{sec:appendix_munrong}

There are two notable difference compared to typical topic models, including Sup-TM-C model. Whereas in most cases specification of one distribution $\boldsymbol\beta_k$ for each type $k$, here $\boldsymbol\beta_k$ denotes a collection of $J$ separate type distributions, $\beta_k^j$, that allow each question to have a unique feature space. Second, while topic models typically allow different words in the same document to have different latent topic assignments, each individual in this model is assigned a single type which then determines the distributions over all questions.
The specific data-generating process for the dynamic survey response (DSR) model is the following:
\begin{equation}
    \tag{DSR}
    \label{eq:d-lda-s}
    \begin{aligned}
    \boldsymbol\beta^j_k &\sim \text{Dirichlet}(\eta^j) \\
    \sigma_k &\sim \text{InverseGamma}(v_0, s_0)\\
    \tilde\theta_{k, t} &\sim \text{Normal}(\tilde\theta_{k,  t-1}, \sigma_k^2)\\
    \boldsymbol\theta_t &= \text{Softmax}(\tilde{\boldsymbol{\theta}_t}) \\
    z_{d,t} &\sim \text{Multinomial}(\boldsymbol\theta_t)\\
    x_{d,j,t} &\sim \text{Multinomial}(\beta^j_{z_{d,t}})\\
    \end{aligned}
\end{equation}

Graphically, it can be represented as the following plate diagram:

\begin{figure}[htb]
    \caption{Plate Diagram for the Dynamic Survey Response Model.}
    \label{fig:DynamicLDAplate}
    \centering
    \begin{tikzpicture}

    \node[circle, draw, inner sep=0pt, minimum size=1cm, fill=lightgray] (xij) at (-4, -2) {$x_{d,j,t}$};
    \node[circle, draw, inner sep=0pt, minimum size=1cm, fill=lightgray] (xijtp1) at (0, -2) {$x_{d,j,t+1}$};
    \node[circle, draw, inner sep=0pt, minimum size=1cm] (beta) at (-2, -2) {$\bm\beta_k$};
    \node[circle, draw, inner sep=0pt, minimum size=1cm] (zi) at (-4, 0) {$z_{d,t}$};
    \node[circle, draw, inner sep=0pt, minimum size=1cm] (zitp1) at (0, 0) {$z_{d,t+1}$};
    \node[circle, draw, inner sep=0pt, minimum size=1cm] (pi) at (-4, 2) {$\bm\theta_{t}$};
    \node[circle, draw, inner sep=0pt, minimum size=1cm] (pitp1) at (0, 2) {$\bm\theta_{t + 1}$};

    \node[circle, draw, inner sep=0pt, minimum size=1cm] (sigma) at (-2, 4) {$\sigma$};

    \draw[->, >=stealth] (beta.west) -- (xij.east);
    \draw[->, >=stealth] (beta.east) -- (xijtp1.west);

    \draw[->, >=stealth] (zi.south) -- (xij.north);
    \draw[->, >=stealth] (zitp1.south) -- (xijtp1.north);

    \draw[->, >=stealth] (pi.south) -- (zi.north);
    \draw[->, >=stealth] (pitp1.south) -- (zitp1.north);
    \draw[->, >=stealth] (pi.east) -- (pitp1.west);
    \draw[->, >=stealth] (sigma.south) -- (pi.north);
    \draw[->, >=stealth] (sigma.south) -- (pitp1.north);

    \draw[->, >=stealth] (-5, 2) -- (pi.west) ;
    \draw[->, >=stealth] (pitp1.east) -- (1, 2);

    \draw (-4.8, -3.8) rectangle (-3.2, 0.8);
    \draw (-0.8, -3.8) rectangle (0.8, 0.8);

    \draw (-2.7, -3.2) rectangle (-1.3, -1.2);

    \draw (-4.7, -3.2) rectangle (-3.3, -1.2);
    \draw (-0.7, -3.2) rectangle (0.7, -1.2);

    \node at (-3.6, -3.5) {$N_t$};
    \node at (.4, -3.5) {$N_{t+1}$};

    \node at (-3.6, -2.9) {$J$};
    \node at (.4, -2.9) {$J$};
    \node at (-1.5, -2.9) {$K$};

    \end{tikzpicture}

    \footnotesize
    \textbf{Note:} The figure presents a simplified diagram of two periods of the DSR model. The number of topics/types $K$ and survey questions $J$ are fixed, while the number of respondents $N_t$ can vary. The rectangles display conditionally independent observations and latent variables. The shaded circles represent observed data, while the white circles represent (unobserved) latent variables. Prior distributions on latent variables are omitted for simplicity.
\end{figure}

In earlier draft of the paper we reported on simulation exercise that showed that inference based on Hamiltonian Monte Carlo, implemented with Numpyro, significantly outperforms the original inference algorithm (stochastic gradient Langevin dynamics) implemented in the original paper, both in terms of computation time and inference quality. These results and associated code are available on request. 